\documentclass[doublecol]{epl2} 
\newcommand{\ie}{{\em i.e. }}
\newcommand{\eg}{{\em e.g. }}
\usepackage{amsmath}
\usepackage{qsymbols}

\title{Role of friction-induced torque in stick-slip motion}
\shorttitle{} 

\author{J. Scheibert\inst{1} \and D.K. Dysthe\inst{1}}
\shortauthor{J. Scheibert \etal}

\institute{                    
  \inst{1} Physics of Geological Processes, University of Oslo, P.O. Box 1048 Blindern, 0316 Oslo, Norway
  }

\abstract{We present a minimal quasistatic 1D model describing the kinematics of the transition from static friction to stick-slip motion of a linear elastic block on a rigid plane. We show how the kinematics of both the precursors to frictional sliding and the periodic stick-slip motion are controlled by the amount of friction-induced torque at the interface. Our model provides a general framework to understand and relate a series of recent experimental observations, in particular the nucleation location of micro-slip instabilities and the build up of an asymmetric field of real contact area.}

\begin{document}

\maketitle

Interfacial friction \cite{Persson-Springer-2000, Urbakh-Klafter-Gourdon-Israelachvili-Nature-2004, Baumberger-Caroli-AdvPhys-2006, Bhushan-Springer-2008} plays a major role in seismology \cite{Scholz-CUP-2002}, biology \cite{Urbakh-Klafter-Gourdon-Israelachvili-Nature-2004, Scheibert-Leurent-Prevost-Debregeas-Science-2009} and nanomechanics \cite{Bhushan-Springer-2008}. The frictional behavior of a contact interface is controlled by the shear strength field $\sigma_c (\bf{x})$, with $\bf{x}$ the position in the interface. As the shear force is increased, a slip region nucleates at the first point where the shear stress reaches the shear strength and grows through the propagation of a micro-slip front. Macroscopic sliding starts only after the entire interface has slipped. This general picture has provided the basis for friction models for decades \cite{Burridge-Knopoff-BSSA-1967, Carlson-Langer-PhysRevLett-1989, Olami-Feder-Christensen-PhysRevLett-1992, Kanamori-Brodsky-RepProgPhys-2004, Brener-Malinin-Marchenko-EurPhysJE-2005, Brochard-deGennes-EurPhysJE-2007, Braun-Barel-Urbakh-PhysRevLett-2009}. Recently, transitions from static to kinetic friction received renewed interest due to experiments that allow the local dynamics of frictional interfaces to be directly measured (see e.g. \cite{Rosakis-Samudrala-Cocker-Science-1999, Rubinstein-Cohen-Fineberg-Nature-2004, Scheibert-Debregeas-Prevost-CondMat-2008, Chateauminois-Fretigny-Olanier-PhysRevE-2010}). Depending on the contact configuration, different kinds of transitions are observed.

\textit{For contacts between bodies having different shapes} (e.g. a sphere on a plane) the transition is smooth. As the shear force is increased, micro-slip occurs immediately at the periphery of the contact where the pressure vanishes, and the slip zone quasistatically invades the higher pressure central region \cite{Scheibert-Debregeas-Prevost-CondMat-2008, Chateauminois-Fretigny-Olanier-PhysRevE-2010}. For multicontacts (contacts between rough solids), this behavior was predicted decades ago by Cattaneo and Mindlin using Amontons' law of friction ($\sigma_c (\bf{x})$=$\mu p(\bf{x})$ where $p$ is the pressure and $\mu$ is the friction coefficient), as follows. The distribution of pressure $p(\bf{x})$ and shear stress $\sigma(\bf{x})$ for a non-slipping interface is first calculated. In an annular region at the contact's periphery, $\sigma (\bf{x}) > \sigma_c (\bf{x})$, showing that slip must take place. Such knowledge of the geometry of the slip region allows the stress field for partial slip conditions to be calculated \cite{Johnson-CUP-1985}. In the following, we will apply a similar procedure to planar contacts.

\textit{For planar contacts}, $\sigma_c (\bf{x})$ is expected to be essentially homogeneous and micro-slip nucleation should occur at random locations due to unavoidable heterogeneities. Surprisingly, in most experiments, and whatever the way the normal and shear loads are applied, micro-slip starts at the trailing edge of the contact and propagates dynamically towards the leading edge \cite{Baumberger-Caroli-Ronsin-PhysRevLett-2002, Baumberger-Caroli-Ronsin-EurPhysJE-2003, Rubinstein-Cohen-Fineberg-Nature-2004, Rubinstein-Cohen-Fineberg-PhysRevLett-2007, Bennewitz-etal-JPhysCondensMatter-2008, Rubinstein-Cohen-Fineberg-JPhysD-2009, Maegawa-Suzuki-Nakano-TribolLett-2010}. Recent experiments on multicontact interfaces also show that the onset of sliding is preceded by a series of precursors of increasing length, which arrest before reaching the interface's leading edge \cite{Rubinstein-Cohen-Fineberg-PhysRevLett-2007, Rubinstein-Cohen-Fineberg-JPhysD-2009, Maegawa-Suzuki-Nakano-TribolLett-2010}. These precursors have also been observed indirectly in microstructured contacts \cite{Bennewitz-etal-JPhysCondensMatter-2008}.

Two aspects of these findings triggered an active debate. One concerns the \textit{dynamics} of the three types of micro-slip fronts observed in \cite{Rubinstein-Cohen-Fineberg-Nature-2004, Rubinstein-Cohen-Fineberg-PhysRevLett-2007, Rubinstein-Cohen-Fineberg-JPhysD-2009}. This variety has recently been reproduced in a 1D spring-block model in which the interface obeys a complex dynamics described by arrays of springs having a distribution of detachment force thresholds and a constant delay time for reattachement \cite{Braun-Barel-Urbakh-PhysRevLett-2009}. This first aspect will not be further addressed here.

The second aspect, which is the main focus of this Letter, concerns the \textit{kinematics} of the precursors, i.e. how their triggering force, number and length are selected. Precursors are always found to nucleate near the trailing edge. They are accompanied by the growth of an asymmetric field of real contact area, with a minimum near the trailing edge, which is retained during macroscopic motion \cite{Rubinstein-Cohen-Fineberg-PhysRevLett-2007, Rubinstein-Cohen-Fineberg-JPhysD-2009}. Conversely, changing the way the shear \cite{Rubinstein-Cohen-Fineberg-PhysRevLett-2007, Rubinstein-Cohen-Fineberg-JPhysD-2009} or normal \cite{Maegawa-Suzuki-Nakano-TribolLett-2010} load is applied modifies the series of precursors. To date, no general description includes all these kinematic features. However, these observations strongly indicate that precursors are highly dependent on how both the pressure and shear stress are distributed at the interface. 

Experiments on planar contacts can be sorted into two groups according to the choice of macroscopic loading: the slider is either pushed from its trailing edge side \cite{Rubinstein-Cohen-Fineberg-Nature-2004, Rubinstein-Cohen-Fineberg-PhysRevLett-2007, Rubinstein-Cohen-Fineberg-JPhysD-2009, Maegawa-Suzuki-Nakano-TribolLett-2010} or driven from its top by a rigid body \cite{Baumberger-Caroli-Ronsin-PhysRevLett-2002, Baumberger-Caroli-Ronsin-EurPhysJE-2003, Bennewitz-etal-JPhysCondensMatter-2008}. In the first configuration, it has been suggested that slip must nucleate at the trailing edge because pushing dramatically increases the shear stress in its vicinity \cite{Rubinstein-Cohen-Fineberg-JPhysD-2009, Braun-Barel-Urbakh-PhysRevLett-2009}. In the second configuration, the shear stress is expected to be distributed homogeneously over the interface, and the reason why nucleation also occurs at the trailing edge is still unclear. In the following, we will focus on this second, top-driven, configuration because the description of the stresses is much simpler than for the first (which would in particular include the contact stress field around the pushing point).

In this Letter, we present a minimal 1D quasistatic model for a sheared planar frictional interface. It involves a stress analysis which is inspired by that of Cattaneo and Mindlin. By accounting for the first time explicitly for the torque that arises whenever the tangential force is not applied exactly in the plane of the interface, we reproduce most of the above-mentioned kinematic observations. This provides a comprehensive picture of the kinematics of the transition from static friction to periodic stick-slip motion, including precursors.

\begin{figure}[!t]
\begin{center}
\includegraphics[width=1.0\columnwidth]{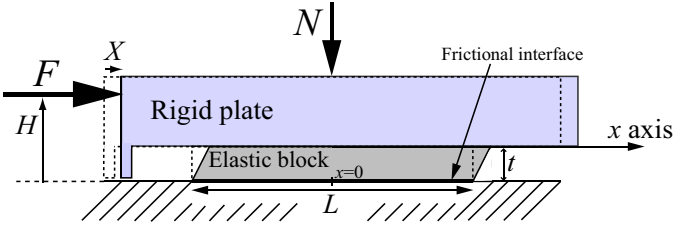}
\caption{Sketch of the system geometry (see text for details) for $X=0$ (dashed line) and $X\neq0$ (solid line). \label{sketch}}
\end{center}
\end{figure}

\textit{Model.} We consider the setup sketched in fig.~\ref{sketch}. A frictional interface is formed by pressing a linear elastic block (Young's modulus $E$, thickness $t$, width $w$ and length $L$, perfectly bonded to a rigid plate) on a horizontal plane. The normal load $N$, applied symmetrically with respect to the contact center, induces no tilt of the plate. The tangential displacement of the plate, $X$, is prescribed at a height $H$ with respect to the frictional interface, giving rise to a friction force $F$. We assume Amontons' rigid-plastic law of friction, with a static friction coefficient $\mu_s$ and a kinetic friction coefficient $\mu_d<\mu_s$. We therefore neglect any effect of the tangential stiffness of the multicontact interface, which would induce deviations to Amontons in contacts where stick and slip regions coexist \cite{Scheibert-Prevost-Frelat-Rey-Debregeas-EPL-2008}. We also neglect state and rate effects \cite{Baumberger-Caroli-AdvPhys-2006}. We place the origin of $x$ at the center of the contact, which extends between $\pm \frac{L}{2}$. The problem is made dimensionless by expressing coordinates in units of $L$, forces in units of $\mu_s N$ and stresses in units of $\frac{\mu_s N}{wL}$. Dimensionless quantities bear a tilde. All physical quantities can be expressed in terms of only two dimensionless control parameters: $r=\frac{\mu_d}{\mu_s}$ and $g=6 \mu_d \frac{H}{L}$.

\begin{figure}[!t]
\begin{center}
\includegraphics[width=1.0\columnwidth]{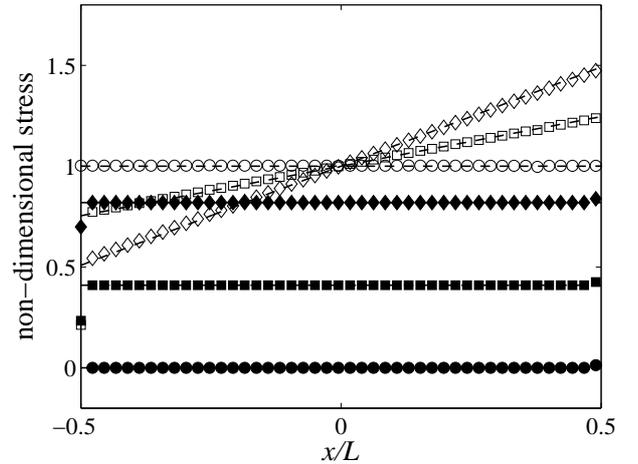}
\caption{Calculated stress fields $\tilde{p}(\tilde{x})$ (open symbols), and $\tilde{\sigma}(\tilde{x})$ (full symbols) for $F$=0 (circles), 0.41$N$ (squares) and 0.82$N$ (diamonds). Stresses are made non-dimensional by dividing them by the average pressure. For clarity, only 1 data point out of 9 is plotted. These fields are compared to the fields assumed in the model (dashed lines).}
\label{fem}
\end{center}
\end{figure}

In order to guide our assumptions about the stress distributions we first performed simple, plane strain, finite element calculations (software Castem 2007) on the configuration sketched in fig. \ref{sketch}. The boundary conditions are the following: no displacement is allowed at the contact interface ; the linear elastic block ($L$=40cm, $t$=2mm, $E$=2.10$^3$Pa, Poisson's ratio 0.4) is perfectly bonded to the rigid plate ($E$=2.10$^{15}$Pa, Poisson's ratio 0.4) ; the normal displacement of the rigid plate is prescribed at the center of its top surface ; the tangential displacement of the rigid plate is prescribed on a point of its left side at a heigth $H$=4cm. We used a regular mesh size of 1mm and QUA4 elements. Figure \ref{fem} shows the pressure and shear stress fields over the contact interface for normal loading only and for two different shear forces applied. For $F$=0, the shear stress is zero and the pressure field is a constant. For $F\neq$0, the shear stress increases homogeneously over the whole contact, whereas the pressure develops an asymmetry which, to a very good approximation, is linear with $x$. We find deviations at both contact extremities due to edge effects, which are significant over distances $\sim t$ for the pressure and $\sim 3t$ for the shear stress. Based on these preliminary calculations, we develop the following model.

\begin{figure*}[!hbt]
\begin{center}
\includegraphics[width=1.0\textwidth]{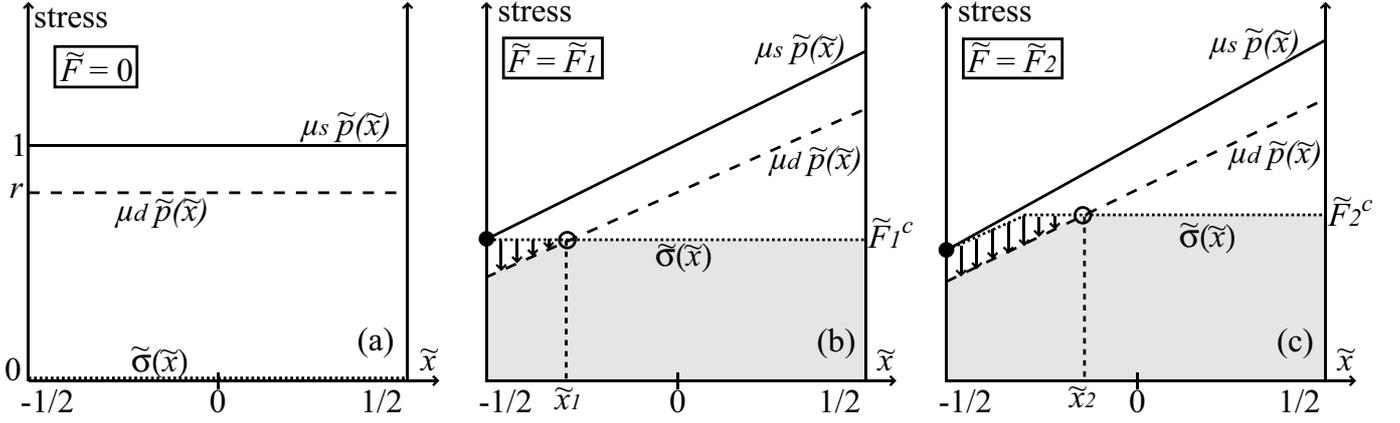}
\caption{Stress fields $\mu_s \tilde{p}(\tilde{x})$ (solid line), $\mu_d \tilde{p}(\tilde{x})$ (dashed line) and $\tilde{\sigma}(\tilde{x})$ (dotted line). (a) $\tilde{F}=0$: all fields are homogeneous and $\tilde{\sigma}(\tilde{x})=0$. (b) $\tilde{F}=\tilde{F}_1$: $\mu_s \tilde{p}=\tilde{\sigma}=\tilde{F}^c_1$ at $\tilde{x}=-1/2$ (black disk). Propagation stops at $\tilde{x_1}$ (open disk) where $\mu_d \tilde{p}=\tilde{\sigma}=\tilde{F}^c_1$. Vertical arrows show the stress relaxation on the segment $[-1/2 ; \tilde{x}_1]$. The grey surface represents the relaxed force $\tilde{F}^*_1$. (c) Same as (b) but for $\tilde{F}=\tilde{F}_2$.
\label{stress}}
\end{center}
\end{figure*}

We neglect any edge effect like \eg a divergence of the stresses at the border of the punch-like contact \cite{Johnson-CUP-1985}, an assumption that is increasingly valid as $t/L$ decreases. These edge effects would be symmetric with respect to $\tilde{x}=0$ and hence could not account alone for the observed asymmetry. Under these conditions, for $\tilde{F}=0$ the pressure field $\tilde{p}(\tilde{x})=\frac{1}{\mu_s}$ is therefore constant (fig.~\ref{stress}(a)). Moreover, between two micro-slip events, the tangential displacement of the top surface of the block is homogeneous, so that the shear stress field $\tilde{\sigma}(\tilde{x})$ increases homogeneously (fig.~\ref{stress}(b)).

For $\tilde{F}\neq 0$, a torque $F H$ is applied at the interface, which can only be balanced by an asymmetry of the pressure field. The elastic block is confined between two rigid planes, one of which (the plate) undergoes a slight tilt. This yields a linear spatial distribution for the normal compression, which is assumed to translate into a linear pressure field $\tilde{p}(\tilde{x},\tilde{F})=\frac{1}{\mu_s} + \frac{2 g}{\mu_d} \tilde{F} \tilde{x}$, which agrees very well with the one calculated above (see fig. \ref{fem}). This assumption is increasingly valid with decreasing $t/L$, and would increasingly break for an increasing compliance of the driving plate. We assume that this linear form remains true even after micro-slips, \ie the pressure field, just like the torque, depends only on the total friction force $\tilde{F}$ and not on the distribution of shear stress over the contact. This is similar to the classical Goodman assumption in contact mechanics \cite{Johnson-CUP-1985, Scheibert-Prevost-Debregeas-Katzav-AddaBedia-JMechPhysSolids-2009}, which is here increasingly valid with decreasing $t/L$.

At a certain force $\tilde{F}=\tilde{F}_1$ the local static slip threshold $\tilde{\sigma}(\tilde{x}) = \mu_s \tilde{p}(\tilde{x},\tilde{F})$ is reached for the first time at the trailing edge $\tilde{x}=-\frac{1}{2}$ (if $g>0$), where the pressure is minimum (black disk in fig.~\ref{stress}(b)), and we find that:
\begin{equation}
\tilde{F}_1=\frac{1}{1+g/r}.\label{F1}
\end{equation}
A micro-slip front nucleates at the trailing edge, turning the problem into the one of an interfacial shear crack. This crack is unstable in the sense of the Griffith energetic criterion (see Appendix) over virtually all $\tilde{x}$ such that $\tilde{\sigma}(\tilde{x}) \in [\mu_d \tilde{p} ; \mu_s \tilde{p}]$. The micro-slip front therefore propagates towards the leading edge. By assuming that the friction force relaxation associated with the micro-slip occurs only after propagation is over, we find that the front arrests at point $\tilde{x}_1$ such that $\tilde{\sigma}(\tilde{x}_1)= \mu_d \tilde{p}(\tilde{x}_1,\tilde{F}_1)$ (open disk in fig.~\ref{stress}(b)).

In all the slipped region, the shear stress drops to $\mu_d \tilde{p}(\tilde{x},\tilde{F}_1)$ (arrows in fig.~ \ref{stress}(b)) so that the shear stress field is now $\tilde{\sigma}(\tilde{x})=r+2 g \tilde{F}_1 \tilde{x}$ for $\tilde{x} \in [-1/2 ; \tilde{x}_1]$
and $\tilde{\sigma}(\tilde{x})=\tilde{F}_1$ elsewhere. We therefore neglect the extension $\sim t$ over which the slope discontinuity of the shear stress at $\tilde{x}_1$ is regularized. The relaxed friction force $\tilde{F}_1^*$ is the integral of this relaxed shear stress field (grey area in fig.~\ref{stress}(b)). The subsequent evolution for increasing $\tilde{F}$ is a series of micro-slip events following the same scenario (see fig.~\ref{stress}(c) for a sketch of the second event). 

We now derive the iteration formulae for the successive values of $\tilde{F}_i$ for such precursors, the corresponding arrest positions $\tilde{x}_i$ and the relaxed forces $\tilde{F}_i^*$. We introduce the elastic force $F^c=K X$, where $K$ is the effective stiffness of the elastic block. $F^c$ is the tangential force that would have been required to move the rigid plate in the absence of partial relaxations related to the micro-slip events. Before macroscopic motion, $\tilde{F}^c$ reduces to $\tilde{\sigma}(\tilde{x}=1/2)$ (see fig.~\ref{stress}). The value $\tilde{F}_i^c$ at the onset of the $i$-th precursor is:
\begin{align}
\tilde{F}_i^c=\tilde{F}_i+\sum^{i-1}_{j=1} \left(\tilde{F}_j - \tilde{F}_j^*\right) \qquad (\tilde{F}_1^c=\tilde{F}_1). \label{Fic}
\end{align}

The arrest point for the $i$-th event is then given by $\mu_d \tilde{p}(\tilde{x}_i,\tilde{F}_i) = \tilde{F}_i^c$ (see fig.~\ref{stress} (b) and (c)), yielding:
\begin{align}
\tilde{x}_i=\left(\tilde{F}_i^c - r\right)/ 2 \tilde{F}_i g.\label{xi}
\end{align}

Knowing $\tilde{F}_i$, $\tilde{F}_i^c$ and $\tilde{x}_i$, the relaxed force $\tilde{F}_i^*$ is derived by integrating the shear stress field just after the precursor. One shows graphically (fig.~\ref{stress}, grey areas) that:
\begin{align}
\tilde{F}_i^*=\tilde{F}_i^c - \frac{1}{2} \left(\tilde{x}_i + {1}/{2}\right) \left[\tilde{F}_i^c - r \left( 1 - {g \tilde{F}_i}/{r}\right)\right] \label{Fi*}
\end{align}
We finally calculate the force at the next event. The static threshold is always reached first at $\tilde{x}=-\frac{1}{2}$. Then $\tilde{F}_{i+1}$  verifies $\mu_d \tilde{p}(-\frac{1}{2},\tilde{F}_i)+\tilde{F}_{i+1}-\tilde{F}_i^*=\mu_s \tilde{p}(-\frac{1}{2},\tilde{F}_{i+1})$, yielding:
\begin{align}
\tilde{F}_{i+1}=\frac{1 - r + \tilde{F}_i^* + g \tilde{F}_i}{1 + {g}/{r}}.\label{Fi+1}
\end{align}
The evolution of the system can be solved iteratively with Eq.~\ref{F1} as the starting point. At each step, $\tilde{F}_i^*$, $\tilde{F}_{i+1}$, $\tilde{F}_{i+1}^c$ and $\tilde{x}_{i+1}$ are computed successively. This procedure has to be slightly modified as soon as $\tilde{x}_{n+1} > 1 / 2$, i.e. when the $(n+1)$-th event first reaches the leading edge after $n$ precursors. At this point, the shear stress relaxes to $\tilde{\sigma}(\tilde{x})=\mu_d \tilde{p}(\tilde{x})$ everywhere, \ie $\tilde{F}_{n+1}^* = r$. The subsequent evolution is then obtained by simply replacing  Eq.~\ref{xi} by $\tilde{x}_i=1/2$ and Eq.~\ref{Fi*} by $\tilde{F}_i^* = r$ in the iteration. 

\begin{figure}[!t]
\begin{center}
\includegraphics[width=1.0\columnwidth]{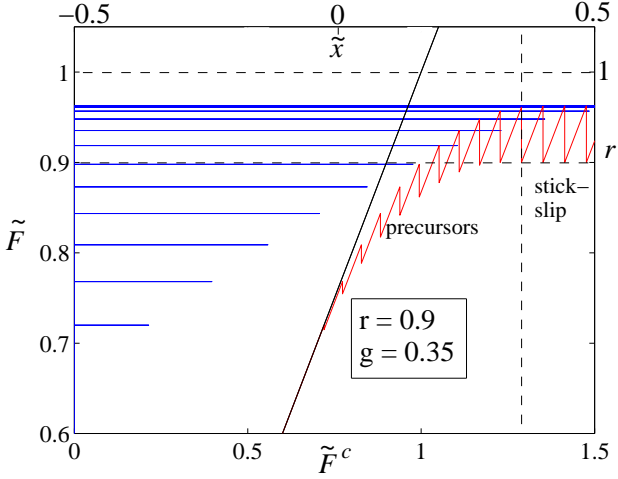}
\caption{Red : loading curve $\tilde{F}(\tilde{F}^c)$ ($r=0.9$, $g=0.35$). Black: ideal elastic loading. Blue horizontal lines: extension of the successive precursors along the contact. \label{charge}}
\end{center}
\end{figure}

\textit{Results.} Figure \ref{charge} shows how the friction force $\tilde{F}$ typically evolves with the elastic force $\tilde{F}^c$ (which is proportional to the prescribed displacement $X$). This loading curve exhibits an initial elastic regime at low forces. It is followed by a series of $n$ relaxations corresponding to $n$ precursors of increasing length. The system then enters a macroscopic stick-slip motion regime (each slip event involves the whole interface), in which the relaxed force is always $r$. Stick-slip progressively becomes periodic, reaching a maximum friction force of $\tilde{F}_{max}=\frac{1}{1 + \frac{g}{r} \left(1 - r\right)}$ (calculated by imposing $\tilde{F}_{i+1}=\tilde{F}_i$ in Eq. \ref{Fi+1}).

\begin{figure}[!t]
\begin{center}
\includegraphics[width=1.0\columnwidth]{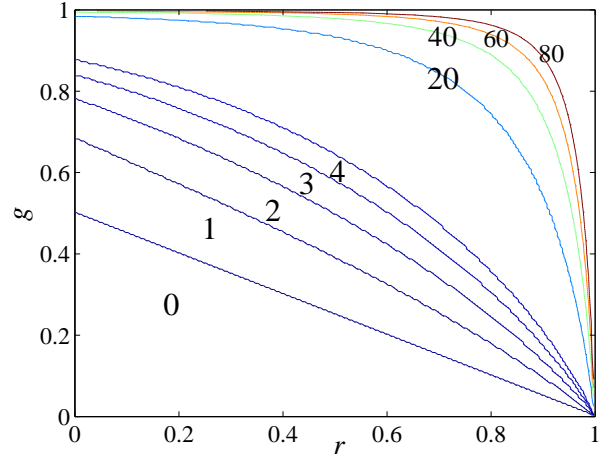}
\caption{Number of precursors as a function of $r$ and $g$.\label{number}}
\end{center}
\end{figure}

In practice, the two control parameters can only be varied in the range $0<r<1$ ($r>1$ is not physical because $\mu_s > \mu_d$) and $0<g<1$ ($\tilde{p}>0$ everywhere in the contact in the absence of adhesive forces, which translates into the condition that, for all $\tilde{F}<\tilde{F}_{max}$, $\tilde{p}(-1/2,\tilde{F})>0$, yielding $r - g \tilde{F}_{max} > 0$ and eventually $g<1$). Figure \ref{number} shows the number of precursors $n$ over the whole accessible parameter space. $n$ increases with both $r$ and $g$. In particular, the larger the torque, the more precursors. Below the line $g=(1-r)/2$, the extension of the first precursor $\tilde{x}_1$ is longer than the contact size, so that no precursor is observed. In particular, no precursor can occur at a torque-free interface ($H=0$, i.e. $g=0$).

\textit{Discussion.} We built a \textit{minimal} model accounting for the rich kinematics of the transition from static to kinetic friction. The key ingredient is the increasing asymmetry induced by the growing tangential force, as soon as the latter is not applied exactly in the plane of the contact interface. We emphasize that this effect is different in essence from any time-invariant pressure asymmetry that would be due to an asymmetric normal loading, like in \cite{Maegawa-Suzuki-Nakano-TribolLett-2010}. Our strongest assumption is that the micro-slip front is so fast that the associated force relaxation occurs only after its arrest. This is a hidden assumption about the propagation dynamics, which is otherwise explicitly beyond the scope of this quasistatic model. In this respect, our model is a crude description of a real frictional interface. Yet, we believe that it captures important generic aspects of sheared frictional interfaces. 

An important component of the present model is the build up of a friction-induced torque. We emphasize that in typical situations, a significant torque corresponds only to a very slight tilt angle. For instance, for small deformations $F/wLE<10^{-3}$, a force applied at the bottom of the rigid plate $H = t$ and an aspect ratio $t/L < 10^{-1}$ yield $\alpha <1.2\cdot10^{-4}$ rad. The asymmetry of the contact is therefore very difficult to avoid experimentally.

Even for an initially homogeneous interface, tangential loading produces an increasing contact pressure asymmetry. In practice, the tangential force is usually applied above the interface ($H>0$), making the pressure lower at the trailing edge. Under these conditions, any friction law that prescribes a monotonic increase in shear strength with increasing local pressure implies that the lowest shear strength is at the trailing edge. In practice most friction laws belong to this category. Here, we considered the classical Amontons' law, which is known to be valid for macroscopic multicontact interfaces \cite{Persson-Springer-2000, Persson-Sivebaek-Samoilov-Khao-Volokitin-Zhang-JPhysCondensMatter-2008}. The smooth glass-on-gel interfaces used in \cite{Baumberger-Caroli-Ronsin-PhysRevLett-2002, Baumberger-Caroli-Ronsin-EurPhysJE-2003}, although a very different system, also obey a pressure-increasing shear strength law, which is a modified version of Amontons' law including adhesive forces. The pressure dependence of shear strength of the microstructured contacts studied in \cite{Bennewitz-etal-JPhysCondensMatter-2008} was not reported but we speculate that it also follows an increasing trend. Our model therefore successfully explains why micro-slip fronts always occur at the trailing edge in both systems \cite{Baumberger-Caroli-Ronsin-PhysRevLett-2002, Baumberger-Caroli-Ronsin-EurPhysJE-2003, Bennewitz-etal-JPhysCondensMatter-2008}, which are top-driven. 

In the side-pushed multicontacts studied in \cite{Rubinstein-Cohen-Fineberg-PhysRevLett-2007, Rubinstein-Cohen-Fineberg-JPhysD-2009, Maegawa-Suzuki-Nakano-TribolLett-2010}, we also expect the torque to decrease the pressure at the trailing edge. This is confirmed by measurements
of the distribution of real contact area just after the successive precursors and after each slip in the periodic stick-slip regime \cite{Rubinstein-Cohen-Fineberg-PhysRevLett-2007, Rubinstein-Cohen-Fineberg-JPhysD-2009}. In a multicontact interface, the real contact area is well-known to be proportional to the applied contact pressure \cite{Greenwood-Williamson-ProcRSocLondA-1966, Persson-Sivebaek-Samoilov-Khao-Volokitin-Zhang-JPhysCondensMatter-2008}. Our model is therefore consistent with their observation of a real contact area asymmetry which increases with $F$, and which is retained and stable in the macroscopic stick-slip regime, with a minimum near the trailing edge. We emphasize that further quantitative comparison between our model and the measurements for the number and length of the precursors in \cite{Rubinstein-Cohen-Fineberg-PhysRevLett-2007, Rubinstein-Cohen-Fineberg-JPhysD-2009, Maegawa-Suzuki-Nakano-TribolLett-2010} is not possible because the two loading systems are very different. In particular, we believe that the huge shear stress increase near the pushing point will likely play the primary role in reducing the shear strength at the trailing edge, and dominate the torque effect. We also emphasize that, even in top-driven systems \cite{Baumberger-Caroli-Ronsin-PhysRevLett-2002, Baumberger-Caroli-Ronsin-EurPhysJE-2003, Bennewitz-etal-JPhysCondensMatter-2008}, quantitative comparison would require knowledge of the effective height $H$ of the applied friction force, which is seldom provided in the literature. We therefore urge authors to provide, in the future, all the details about the loading configuration that are necessary to assess the level of torque, like in \cite{Bureau-Baumberger-Caroli-PhysRevE-2000} where $H=0$.

An implication of the fact that micro-slip very generically nucleates at the trailing edge of a contact implies that the interface will be in a compressed state as soon as the first precursor occurs. This strongly suggests that the results of Bennewitz \textit{et al} \cite{Bennewitz-etal-JPhysCondensMatter-2008}, which measure such compression, are valid over a much broader range of systems than their microstructured PDMS on glass interface.

Other results of the model are of general interest for frictional systems. First, they suggest that, from a kinematic point of view, macroscopic stick-slip is not different from the preceding precursors, the latter being merely defined by their limited extent over the interface. Second, we show that $F_{max}/N$ is always smaller than $\mu_s$, \ie the macroscopic friction coefficient is always lower than the local friction coefficient, which governs the onset of the micro-slip instability. The macroscopic friction coefficient is moreover dependent on $g$, \ie on the details of the loading system, suggesting that care has to be taken when a friction coefficient is measured using a macroscopic experiment.

To conclude, we have shown that in virtually all friction experiments, slip will occur at an asymmetrically loaded interface due to a friction-induced torque. In particular, the pressure is minimum near the trailing edge, where all micro-slip instabilities nucleate. In top-driven systems, like the one studied here, the \textit{kinematics} of the transition from static to stick-slip friction is then dominated by this macroscopic, system-dependent, asymmetry rather than by small scale heterogeneities of the shear strength. This is in striking contrast with most friction models, which assume a homogeneously loaded interface bearing a relatively small disorder (see \eg \cite{Burridge-Knopoff-BSSA-1967, Carlson-Langer-PhysRevLett-1989, Olami-Feder-Christensen-PhysRevLett-1992, Braun-Barel-Urbakh-PhysRevLett-2009}). In this respect, the present work provides a framework for the extension of these models to take into account a macroscopic shear strength asymmetry.

\section{Appendix}
Here we show, using an energetic criterion, that a crack tip at the frictional interface is unstable at virtually all points $x$ satisfying the following stress criterion $\sigma(x) > \mu_d p(x)$. We apply the Griffith energetic criterion which states that an existing crack will propagate if $G(x) > \Gamma_c(x)$, where $G(x)$ is the energy release rate at point $x$ and $\Gamma_c(x)$ is the facture energy at point $x$. To do this, one needs to regularize Amontons' law of friction by describing how the shear stress drops from $\mu_s$ to $\mu_d$. The easiest way, which is classically used in dynamic crack simulations (see \eg \cite{Scholz-CUP-2002, DiCarli-FrancoisHolden-Peyrat-Madariaga-JGeophysRes-2010}), is to assume a linear slip weakening: $\sigma$ drops linearly from $\mu_s p$ to $\mu_d p$ over a critical weakening distance $D_c$. Then $\Gamma_c (x) = \frac{(\mu_s - \mu_d) D_c}{2} p(x)$.

When slip occurs at $x$, the shear stress is relaxed by the amount $(\mu (x) - \mu_d)p(x)$, where $\mu (x) = \sigma(x)/p(x)$. The amplitude of the corresponding slip is $\delta (x)=t (\mu (x) - \mu_d)\frac{p(x)}{E}$. Neglecting the variation of stress during the first slip over the (microscopic) length $D_c$, the stress during all the slip is $\mu_d p(x)$. Therefore $G(x) \simeq \mu_d (\mu (x) - \mu_d) \frac{t}{E} p^2(x)$.

The condition $G>\Gamma_c$ therefore reads
\begin{equation}
\mu (x) - \mu_d \gtrsim \frac{E}{p(x)} \frac{D_c}{t} \frac{\mu_s - \mu_d}{2 \mu_d}.
\end{equation}

$D_c$ and $t$ being respectively a microscopic and a macroscopic length scale, $D_c/t<<1$. This means that the crack will stop only for very low values of $\mu (x) - \mu_d$. We therefore consider that a precursor practically propagates until the point where $\sigma(x)=\mu_d p(x)$.


\acknowledgments
We are very grateful to E. Barthel, G. Debr\'egeas, A. Malthe-S{\o}renssen, P. Meakin and F. Renard for discussions. We acknowledge funding from the European Union (Marie Curie grant PIEF-GA-2009-237089). This paper was supported by a Center of Excellence grant to PGP from the Norwegian Research Council.

\end{document}